\documentclass[12pt]{article}
\usepackage{epsf}  
\newcommand{\beq }{\begin{eqnarray}}
\newcommand{\eeq }{\end{eqnarray}}
\newcommand{\lr }{\left(}
\newcommand{\rl }{\right)}
\newcommand{\lan }{\langle}
\newcommand{\ran }{\rangle}
\newcommand{\SRa}{ 
\put(0,6) { \vector(1,0){10}  }
\put(10,0){ \vector(-1,0){10} } 
\qquad}
\newcommand{\SRb}{ 
\put(0,10) { \vector(1,-1){10}  }
\put(10,10){ \vector(-1,-1){10} } 
\qquad}

\begin{document}

\title{Spectral Form Factor for Chaotic Dynamics \\ 
in a Weak Magnetic Field}

\author{Keiji Saito {\footnote {\tt email: 
saitoh@spin.phys.s.u-tokyo.ac.jp}}~~and 
Taro Nagao {\footnote {\tt email: 
nagao@math.nagoya-u.ac.jp}}  
}
\date{}
\maketitle

\begin{center}
\it
$^*$ Department of Physics, Graduate School of Science,
University of Tokyo, Hongo 7-3-1, Bunkyo-ku, Tokyo 113-0033, Japan
\end{center}
\begin{center}
\it
$^\dag$ Graduate School of Mathematics,
Nagoya University, Chikusa-ku, \\ Nagoya 464-8602, Japan
\end{center}

\begin{abstract}
Using semiclassical periodic orbit theory for a chaotic system in a
weak magnetic field, we obtain the form factor predicted 
by Pandey and Mehta's two matrix model up to the third order. The third
order contribution has a peculiar term which exists only in the 
intermediate crossover domain between the GOE (Gaussian Orthogonal 
Ensemble) and the GUE (Gaussian Unitary Ensemble) universality classes. 
The exact expression is obtained by taking account of the contribution 
from encounter regions where orbit loops are connected.  
\end{abstract}


\medskip


\newpage

\section{Introduction}

Universal level statistics of classically chaotic systems 
in agreement with the RMT (Random Matrix Theory) prediction 
has been intensively studied since it was conjectured two 
decades ago\cite{BGS84}.
 
Berry\cite{B85} first evaluated the first order term of the 
spectral form factor using the semiclassical periodic orbit 
theory\cite{GW}. His method is called the diagonal approximation. 
It is applicable to both of the GOE (Gaussian Orthogonal Ensemble) 
and GUE (Gaussian Unitary Ensemble) universality classes.
In the absence of time reversal invariance, the first order 
term is evaluated as a sum over the periodic orbits and agrees 
with the prediction derived from the GUE of random matrices. 
If the system is time reversal invariant, we additionally need 
to sum up over the pairs of the periodic orbits mutually 
time reverse. The result is also in agreement with the RMT 
prediction derived from the GOE.

Recently Sieber and Richter (SR) showed that a family of orbit pairs 
with an encounter gave the first correction to the diagonal approximation
\cite{SR01}. Their idea clarifies what corresponds to the {\em order} 
of each term in the context of the periodic orbit theory. Heusler et al. 
\cite{HMBH04} further generalized SR's idea by using a combinatorial 
analysis and M\"uller et al.\cite{MHBHA04,MHBHA05} obtained the 
full expansion which yields the RMT form factor.

In this paper, we investigate the form factor of chaotic 
systems in a weak magnetic field, extending the method 
of Ref.\cite{HMBH04}. Our aim is to reproduce 
the universal level statistics in the crossover domain 
between the GOE and GUE universality classes. The corresponding 
RMT formula is derived from Pandey and Mehta's two matrix 
model\cite{PM83}. This formula interpolates the form 
factors of GOE and GUE, which correspond 
to chaotic systems without and with a magnetic field, respectively.
Using a quantum graph, the authors previously obtained the second order 
term by employing Sieber and Richter's periodic orbit pairs and 
showed that it agreed with the RMT prediction\cite{NS03}. 
Here we consider the third order term of the 
form factor and clarify a mechanism to yield the RMT
expression. We show that the RMT expression is obtained 
by taking account of the interference of the gauge 
potential in the encounters of periodic orbits. 

Let us explicitly present the RMT prediction of the form factor. 
It is derived from Pandey and Mehta's two matrix 
model, as explained in\cite{NS03}. For small $\tau 
(0\le \tau \le {1\over 2})$, the RMT form factor $K_{\rm RM}(\tau)$ 
with a parameter $\mu$ is written as
\beq
K_{\rm RM} (\tau ) &=& \tau + {1\over 2} 
\int_{1-2\tau }^{1}  dk \, {k \over k + 2\tau }
e^{-\mu(k + \tau)}  
\nonumber \\
&=&  \tau + \tau e^{-\mu} - 2\tau^2 e^{-\mu} 
+ \tau^3 \left( 
{\mu^2 \over 6} e^{-\mu} + 
2 e^{-\mu} \right) + \cdots . \label{pd}
\eeq

The GOE and GUE limits correspond to $\mu\to0$ and 
$\mu\to\infty$, respectively. Up to the second order of 
$\tau$, each term is a monotonic function of the parameter 
$\mu$. However, the third order term in (\ref{pd}) 
includes a peculiar nonmonotonic term $\tau^3 
{\mu^2 \over 6} e^{-\mu}$, which vanishes in both of the GOE 
and GUE limits. Such terms generally appear in higher order 
contributions. We can thus expect that a semiclassical 
analysis of the third order term will offer a basis to 
reproduce the full expansion of $K_{\rm RM}(\tau)$. 

\section{Chaotic System in a Magnetic Field and Periodic Orbit Theory}

We consider a quantum system in a magnetic field and 
assume that the corresponding classical dynamics is chaotic. 
In the semiclassical limit $\hbar \rightarrow 0$, the form factor 
of such a system can be 
formally written in terms of the classical periodic 
orbits\cite{GW};
\beq
K_{\rm po}(\tau ) =
\left\langle
\sum_{\gamma , \gamma'} A_{\gamma}A_{\gamma'}^{\ast} e^{i( S_{\gamma}  
+ \theta_{\gamma})/\hbar }
e^{ -i( S_{\gamma'} + \theta_{\gamma'} )/\hbar }
\delta \left( \tau - {T_{\gamma} + T_{\gamma'} \over 2T_{\rm H}}\right)
\right\rangle ,
\label{fmr}
\eeq
where $S_{\gamma}$, $T_{\gamma}$ and $A_{\gamma}$ are 
the action without gauge potential 
(including the Maslov phase), period and 
dimensionless stability amplitude of the 
periodic orbit $\gamma$. The phase $\theta_{\gamma}$ is 
an effect of the gauge potential in the action along 
the orbit $\gamma$, i.e., 
$\theta_{\gamma}=
 \int_{\gamma} d{\bf x}_{\gamma}    \cdot {\bf a}({\bf x}_{\gamma}) 
= \int_{\gamma} dt \, d{\bf x}_{\gamma}/dt \cdot {\bf a}({\bf x}_{\gamma})$, 
using the gauge potential ${\bf a}({\bf x}_{\gamma})$ at the 
position ${\bf x}_{\gamma}$.
The angular bracket stands for an average over 
the mean energy $(E+E')/2$ and over a time interval much smaller than 
the Heisenberg time $T_{\rm H}$. Since the classical dynamics 
is chaotic, each periodic orbit shows a diffusive 
behavior\cite{TR03,R_text}; successive changes of the velocity 
can be regarded as independent events. Hence, if the time $T$ 
elapsed on an orbit $\gamma$ is sufficiently large, we assume 
that the term $g_{\gamma} (t) \equiv d{\bf x}_{\gamma}/dt 
\cdot {\bf a}({\bf x}_{\gamma}) $ can be replaced by the 
Gaussian white noise satisfying the correlation 
$\lan \lan g_{\gamma} (t) g_{\gamma }(t') \ran \ran 
= 2D \delta (t-t')$. Here a Gaussian average $\lan\lan 
\cdots\ran\ran$ is computed as a functional integral 
\beq 
\label{gaussian}
\lan \lan \ F[g_{\gamma}] \ \ran \ran = \frac{\displaystyle 
\int {\cal D}g_{\gamma} \ {\rm exp}\left[ 
-\frac{1}{4 D} \int_0^T dt (g_{\gamma}(t))^2 
\right] F[g_{\gamma}]} 
{\displaystyle \int {\cal D}g_{\gamma} \  
{\rm exp}\left[ -\frac{1}{4 D} \int_0^T dt (g_{\gamma}(t))^2 
\right]}.
\eeq
Incorporating such a Gaussian average, we obtain a formula 
\beq
K_{\rm po} (\tau ) = 
\left\langle
\sum_{\gamma , \gamma'} A_{\gamma}A_{\gamma'}^{\ast}
e^{i( S_{\gamma}  -S_{\gamma'} )/\hbar }
\lan\lan e^{ i( \theta_{\gamma} - \theta_{\gamma'} )/\hbar} \ran\ran
\delta \left( \tau - {T_{\gamma} + T_{\gamma'} \over 2T_{\rm H}}\right)
\right\rangle . \label{pobs}
\eeq
We start from Berry's diagonal approximation, which yields 
the first order term\cite{B85}. This term is obtained from 
two types of periodic orbit pairs 
$({\gamma,\gamma})$ and $({\gamma, \overline{\gamma}})$. Here 
a periodic orbit $\overline{\gamma}$ is the time reverse of $\gamma$. 
For the pairs of identical orbits $({\gamma,\gamma})$, the dependence 
on the gauge potential is canceled, so that the result does not 
depend on the magnetic field. The estimate for the form 
factor is given by the Hannay and Ozorio de Almeida (HOdA)'s 
sum rule\cite{HODA84};
\beq
\sum_{\gamma}\left| A_{\gamma}\right|^2
\delta \lr \tau - {T_{\gamma}\over T_{H}}\rl 
= \tau.
\eeq 
For the pairs of mutually time reverse orbits $({\gamma,\overline{\gamma}})$, 
on the other hand, the result depends on the magnetic field. 
It is estimated as 
\beq
\sum_{\gamma}\left| A_{\gamma}\right|^2
\delta \lr \tau - {T_{\gamma}\over T_{H}}\rl 
\lan\lan e^{ i( \theta_{\gamma} - \theta_{\overline{\gamma}})/\hbar} \ran\ran
=\tau \lan\lan e^{2i \theta_{\gamma} /\hbar } \ran\ran 
= \tau e^{-bT},
\eeq
where the last equality results from the 
Gaussian average (\ref{gaussian}). 
Here $T$ is the period $T(=\tau T_{H})$ and $b$ 
is defined in terms of the diffusion constant $D$ 
as $b = 4D/\hbar^2$, which depends on the magnetic field. 
Putting the above results together, we reproduce the 
first order term of the form factor 
\cite{BGOS95} as
\beq
K_{\rm po}^{(1)} &=& \tau + \tau e^{-bT},
\eeq
which reproduces the first two terms of the 
RMT prediction (\ref{pd}). The RMT parameter 
$\mu$ is identified with $b T$. 

We next consider the second order term. 
As shown by Sieber and Richter (SR) \cite{SR01}, a family of orbit pairs 
with one encounter plays a crucial role\cite{NS03,TR03}.
We can symbolically write SR pairs as $\gamma=E L_1 \overline{E} L_2$ and 
$\gamma'=E \overline{L}_1 \overline{E} L_2$ (see Figure 1). Here $E$ and its almost 
time reverse $\overline{E}$ are the orbit segments in the encounter where two 
loops are connected. The loops are denoted as $L_1$ and $L_2$, 
respectively, and $\overline{L}_1$ is the time reverse of $L_1$. 
Let us see the behavior of the periodic orbits $\gamma$ and 
$\gamma'$ in the encounter region. The directions of encounter 
orbits are depicted as $\SRa$ for 
$\gamma$ and $\SRb$ for $\gamma'$. For a system with two 
degrees of freedom, it is convenient to use the 
Poincar\'e section ${\cal P}$ in the phase space 
\cite{HMBH04}. The orbit $\gamma$ pierces through 
$\cal P$ at two phase space points denoted by vectors 
$x_a$ and $x_b$. Let us express 
${\cal T}x_b - x_a$ (${\cal T}$ is the time reversal 
operator) in terms of the pairwise normalized 
vectors $\hat{e}_{s}$ and $\hat{e}_{u}$ as $
{\cal T}x_b - x_a = u \hat{e}_{u} + s \hat{e}_{s}$.
To demand that the orbits in the encounter are mutually 
close, we introduce a bound $c$ and assume that $|u|,|s| 
< c$. Then the partner orbit $\gamma'$ (or the 
time reverse of $\gamma'$) pierces at $x_a^p = 
x_a + u \hat{e}_{u}$ and at $x_b^p= {\cal T}(x_a 
+ s\hat{e}_{s})$. We can consequently estimate 
the duration $t_{\rm enc}$ of the encounter 
and the action difference $\Delta S$ for the 
two orbits as
\beq
t_{\rm enc} = 
{1\over \lambda} \ln {c^{2}\over |us|} 
\qquad {\rm and}\qquad 
\Delta S = us, 
\eeq
where $\lambda$ is the Liapunov exponent. 

We need to estimate the number of encounters 
in one periodic orbit of a period $T$. 
This can be computed as
\beq
\int_{-c}^c du ds 
\int_{0}^{T -2t_{\rm enc}} d T_1
{T \over {\cal N} t_{\rm enc} \Omega},
\label{enb} 
\eeq
where $\Omega^{-1} (={1\over 2\pi\hbar T_H})$ 
is the ergodic return probability per a unit action. 
The factor $T \int_{0}^{T - 2 t_{\rm enc}} {\rm d}T_1$ 
indicates that one of the two piercings occurs in the 
time interval $[0,T]$ and that the time $T_1$ elapsed 
on the loop $L_1$ or $\overline{L}_1$ lies in $[0,T-2 t_{\rm enc}]$.
The combinatorial factor ${\cal N} = 2$ is 
necessary to avoid a double counting due to 
the equivalence of $E$ and $\overline{E}$ in 
$\gamma$. 

Let us now evaluate the contribution of the gauge potential 
to the second order term. The Gaussian average (\ref{gaussian}) 
yields a factor $e^{-b T_1}$ as a contribution from the 
the pair of the loops $L_1$ and $\overline{L}_1$. In the region of the encounter, 
the gauge potential destructively interferes with itself 
and yields no contribution. That is, since mutually 
almost time reversed orbits are combined in the encounter, 
a product of the phase factors $e^{i \theta_{\gamma}}$ 
and $e^{-i \theta_{\gamma^{\prime}}}$ 
gives a factor $1$. Incorporating the factor $2$ due to the 
time reversal of the orbit $\gamma^{\prime}$, we 
obtain the second order term as
\beq
K_{\rm po}^{(2)}(\tau) &=& 2 \sum_{\gamma}\left| A_{\gamma}\right|^2
\delta \lr \tau - {T_{\gamma}\over T_{H}}\rl
\int_{-c}^{c} du ds
\int_{0}^{T -2t_{\rm enc}} d T_1  
{T \over {\cal N} t_{\rm enc} \Omega} 
e^{-b T_{1}} 
e^{i us/\hbar} 
\nonumber \\
&\rightarrow&-2\tau^2 e^{-b T}, 
\eeq
again in agreement with (\ref{pd}). Here we calculated only the 
term independent of $t_{\rm enc}$, as the other terms vanish 
due to appearances of extra factors $\hbar$ or rapid 
oscillations\cite{MHBHA05}.
 
\section{Third Order Term}
We now calculate the third order contribution, following  
the recipe explained in previous section. We will find 
that the calculation is somewhat different from the second 
order term because the interference of the gauge 
potential in the encounter regions gives a 
significant contribution. For the 
periodic orbit pairs which contribute 
to the third order term, Heusler 
et al.\cite{HMBH04} introduced several 
notations (aas, api, ppi, ac and pc). 
In order to take account of the time 
reversal of the orbit $\gamma^{\prime}$, 
we modify their notations by putting 
suffixes (-a and -b) as drawn in 
Figures 2 and 3. 

Let us first consider a general orbit pair 
$\alpha$ with $L$ loops and $V$ encounters. 
Extending the result (\ref{enb}) for 
the second order term, we can compute 
the number of encounters in one periodic 
orbit of a period $T$ as
\beq 
\int d{\bf u}d{\bf s} 
\int_0^{T-t_{\alpha}} dT_1  
\int_0^{T-t_{\alpha}-T_1} dT_2 \cdots 
\int_0^{T-t_{\alpha}-T_1-T_2-\cdots-T_{L-2}} dT_{L-1} 
\ Q_{\alpha}, 
\eeq
where
\beq
Q_{\alpha} = \frac{  T}{{\cal N}_{\alpha} \  
\prod_{v=1}^V t_{{\rm enc},v} \ \Omega^{L-V}}. 
\eeq
The integration measures are $d{\bf u} = \prod_{j=1}^{L-V} 
du_j$ and $d{\bf s} = \prod_{j=1}^{L-V} ds_j$ in terms of 
appropriate phase space coordinates $(u_j,s_j)$. The times elapsed 
on the $L$ loops are depicted as $T_1,T_2,\cdots,T_L$. 
The duration of the $v$-th encounter is written as $t_{{\rm enc},v}$. 
The total duration of the encounters in one periodic orbit 
is $t_{\alpha} = \sum_{v=1}^V l_v t_{{\rm enc},v}$ ($l_v$ 
is the number of the orbit segments contained in the $v$-th 
encounter). A combinatorial factor ${\cal N}_{\alpha}$ 
depends on the structure of the orbit pair. 
\par 
We then consider the contribution of the gauge potential. 
The Gaussian average (\ref{gaussian}) on the loops yields 
a factor $e^{-b T_1} e^{-b T_2} \cdots e^{-b T_K}$, 
where $K$ is the number of the pairs of mutually 
time reversed loops. In Figures, the encounter orbits 
are schematically drawn along with the passing 
directions through the encounter regions. For the $v$-th 
encounter, we introduce the number $n_{{\rm enc},v}$ of 
the orbit segment pairs which cause the 
significant gauge potential contribution. 
Let us fix an arbitrary direction $(+)$ 
of the orbit passing and call the opposite 
direction $(-)$. Then the number 
$n_{{\rm enc},v}$ can be computed as 
\beq
n_{{\rm enc},v} = \frac{1}{2}
\left| \left\{ \#^{(+)}(\gamma)-\#^{(-)}(\gamma) \right\} - 
 \left\{ \#^{(+)}(\gamma') - \#^{(-)}(\gamma') \right\} \right|.
\eeq
Here $\#^{(+)}(\gamma)$ and $\#^{(-)}(\gamma)$ mean the 
numbers of passings of $\gamma$ through the encounter 
in $(+)$ and $(-)$ directions, respectively. We again compute 
the Gaussian average (\ref{gaussian}) to evaluate the effect 
of the gauge potential. The contribution from an encounter 
is then estimated as $\lan \lan  e^{2 i n_{{\rm enc},v} 
\theta_{\gamma}/\hbar} \ran \ran 
= e^{- b n_{{\rm enc},v}^2 t_{{\rm enc},v}}$. 
   
Putting the above results together, we find the 
contribution to the form factor from the orbit 
pair $\alpha$ as 
\beq
& & K_{\rm po,\alpha}(\tau) 
= \tau \int d{\bf u}d{\bf s} 
\nonumber \\ & \times  & 
\int_0^{T-t_{\alpha}} dT_1  
\int_0^{T-t_{\alpha}-T_1} dT_2 \cdots 
\int_0^{T-t_{\alpha}-T_1-T_2-\cdots-T_{L-2}} dT_{L-1} 
Q_{\alpha} R_{\alpha} e^{i \Delta S/\hbar} \nonumber \\
\label{kpo} 
\eeq
with
\beq
R_{\alpha} = e^{-b(T_1 + T_2 + \cdots + T_K)} 
e^{-b(n_{{\rm enc},1}^2 t_{{\rm enc},1}  
 + n_{{\rm enc},2}^2 t_{{\rm enc},2} + \cdots +   
n_{{\rm enc},V}^2 t_{{\rm enc},V})}.
\eeq
This contributes to the terms of order $L-V+1$. 
Here $\tau$ appears due to the HOdA sum rule 
and the action difference is estimated as 
$\Delta S = \sum_{j=1}^{L-V} u_j s_j$. 

Let us go back to the calculation of the third order 
term with $L-V+1=3$. From Figures 2 and 3, 
we can readily see the followings. For $\alpha=$ aas,api 
and ppi, $L=4$, $V=2$ and $t_{\alpha}=2\lr 
t_{\rm enc, 1} + t_{\rm enc, 2}\rl$. On the other hand, 
for $\alpha=$ ac and pc, $L=3$, $V=1$ and 
$t_{\alpha}=3 t_{\rm enc}$. Thus we can 
calculate $R_{\alpha}$ as
\beq
R_{\rm aas-a}
&=& R_{\rm aas-b} = e^{-b\lr T_1 + T_2 \rl} , \\
R_{\rm api-a} &=& R_{\rm api-b} = e^{-b\lr T_1 + T_2  
+ t_{\rm enc, 1}+ t_{\rm enc , 2} \rl} , \\
R_{\rm ppi-a} &=& 1 , \\
R_{\rm ppi-b} &=& 
e^{-b\lr T -t_{\rm ppi}\rl 
-4b\lr t_{\rm enc , 1}+ t_{\rm enc ,2}\rl } , \\
R_{\rm ac-a} &=& e^{-b\lr T_1 + T_2 \rl} , \\
R_{\rm ac-b} &=& e^{-b\lr T_1 + t_{\rm enc} \rl} , \\
R_{\rm pc-a} &=& 1 , \\
R_{\rm pc-b} &=& e^{-b\lr T -t_{pc} \rl - 9 b t_{\rm enc}} . 
\eeq
Moreover the combinatorial factor ${\cal N}_{\alpha}$ 
is evaluated in \cite{HMBH04} as
\beq 
({\cal N}_{\rm aas}, {\cal N}_{\rm api}, {\cal N}_{\rm ppi},
{\cal N}_{\rm ac}, {\cal N}_{\rm pc})=(2,2,4,1,3).
\eeq

We substitute these formulas into (\ref{kpo}) and 
again extract the terms independent of the encounter 
durations $t_{{\rm enc} ,1}$, $t_{{\rm enc} ,2}$ and 
$t_{\rm enc}$. After a straightforward calculation, 
we obtain contributions from the orbit pairs
\beq
K_{\rm po,aas-a}(\tau ) &=& K_{\rm po, aas-b}(\tau ) = 2 \tau^3 e^{-bT}  , \\
K_{\rm po,api-a}(\tau ) &=& K_{\rm po, api-b} (\tau ) = 
\tau^3 \left[ 
-{e^{-bT} -1 \over bT} + {1 + e^{-bT} \over 2}
\right]  , \\
K_{\rm po,ppi-a}(\tau ) &=&  \tau^3 ,\\
K_{\rm po,ppi-b}(\tau ) &=& \tau^3 e^{-bT} \left[ 
{(bT)^2 \over 6} + bT + 1 \right] ,\\
K_{\rm po,ac-a}(\tau ) &=& -3\tau^3 e^{-bT} ,\\
K_{\rm po,ac-b}(\tau ) &=&  \tau^3 \left[ 
-1 + {2\lr e^{-bT} - 1 \rl \over bT}
\right] ,\\
K_{\rm po,pc-a}(\tau ) &=& -\tau^3 , \\
K_{\rm po,pc-b}(\tau ) &=& -\tau^3 e^{-bT} \lr  1 + bT\rl .
\eeq
Summing up these values, we arrive at the third order term of 
the form factor
\beq
K_{po}^{(3)}(\tau ) = \sum_{\alpha}  K_{\rm po, \alpha}(\tau ) =
 \tau^3 \left( 
{(bT)^2 \over 6} e^{-bT} + 
2 e^{-bT} \right),
\eeq
which agrees with the RMT prediction (\ref{pd}). One might think that 
$K_{\rm po,api}$ and $K_{\rm po,ac}$ should disappear in the 
limit $b\to\infty$, because a pair of mutually time reversed 
loops exists. However, as these contributions come from the 
limiting situations that the times elapsed on those loops are 
short, there is no inconsistency. 

In summary, we investigate the level statistics of 
a classically chaotic system in the crossover domain 
between GUE and GOE universality classes. Summing up 
the contributions from the relevant periodic orbit pairs 
of a chaotic system, we find an agreement between the 
semiclassical form factor and the RMT result up 
to the third order. We expect that this agreement 
holds up to any arbitrary order. The periodic orbit pairs 
consist of loops and encounters. The durations of 
the encounters are logarithmically divergent 
in the limit $\hbar \rightarrow 0$ but 
infinitesimally small compared to the orbit 
periods, which are of the order of the 
Heisenberg time $T_H$ ($\sim O(1/\hbar)$ 
for a system with two degrees 
of freedom)\cite{HMBH04}. Nevertheless, 
the contribution from the encounter regions 
is crucial. This mechanism should also be 
a key factor in the derivation of higher 
order terms. It might moreover shed light 
on a similar counting problem\cite{BSW03} 
for a quantum graph.  

\section*{Acknowledgement}

One of the authors (T.N.) is grateful to Dr. Gregory Berkolaiko, 
Dr. Petr Braun, Dr. Sebastian M\"uller and Dr. Holger Schanz 
for valuable discussions.

\newpage

\begin{figure}[!t]
\epsfxsize=14cm
\centerline{\epsfbox{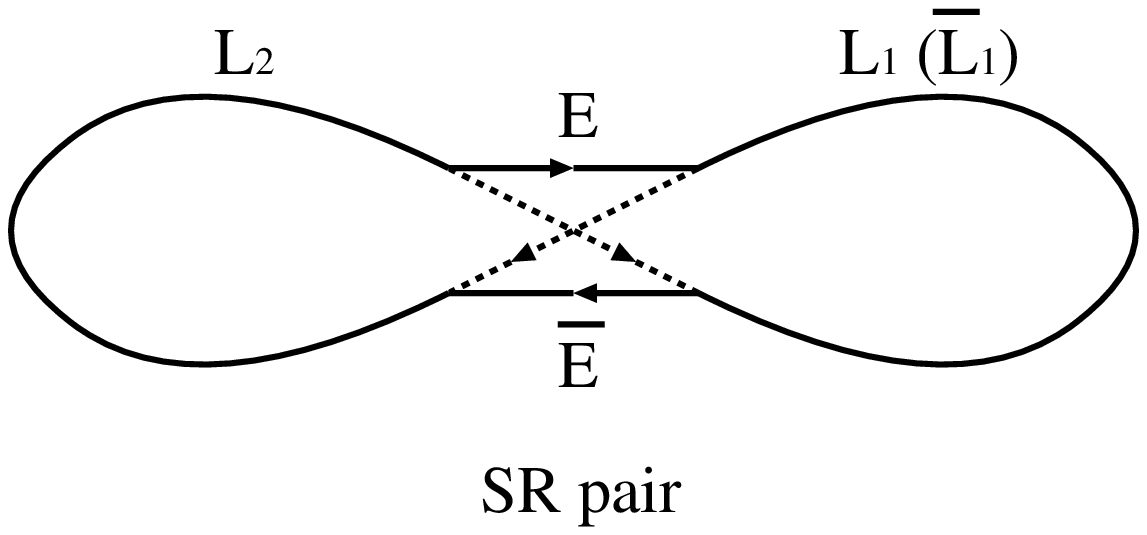}}
\caption{The Sieber-Richter (SR) pair of periodic orbits}
\end{figure}

\begin{figure}[!t]
\epsfxsize=14cm
\centerline{\epsfbox{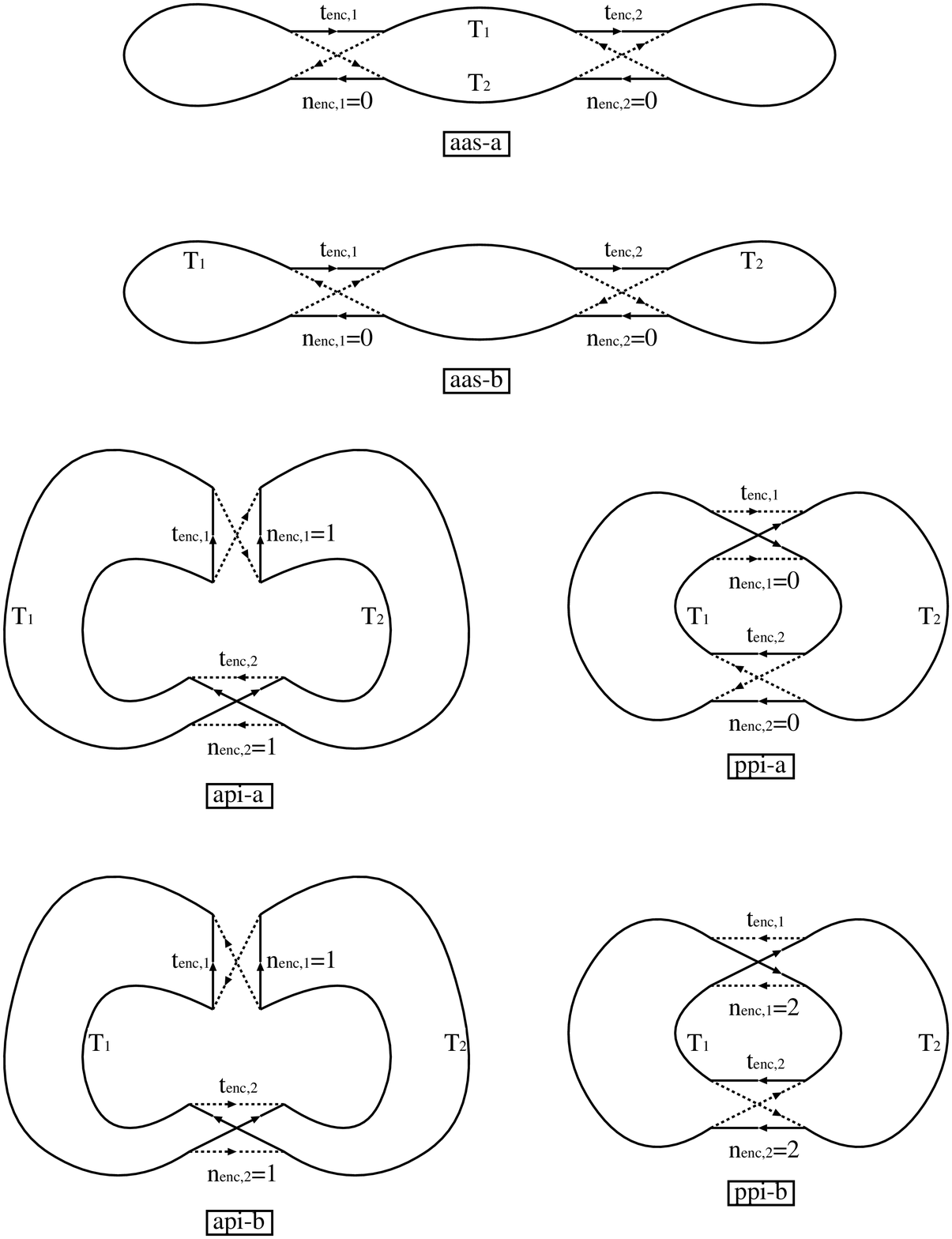}}
\caption{The periodic orbit pairs with two encounters.}
\end{figure}

\begin{figure}[!t]
\epsfxsize=14cm
\centerline{\epsfbox{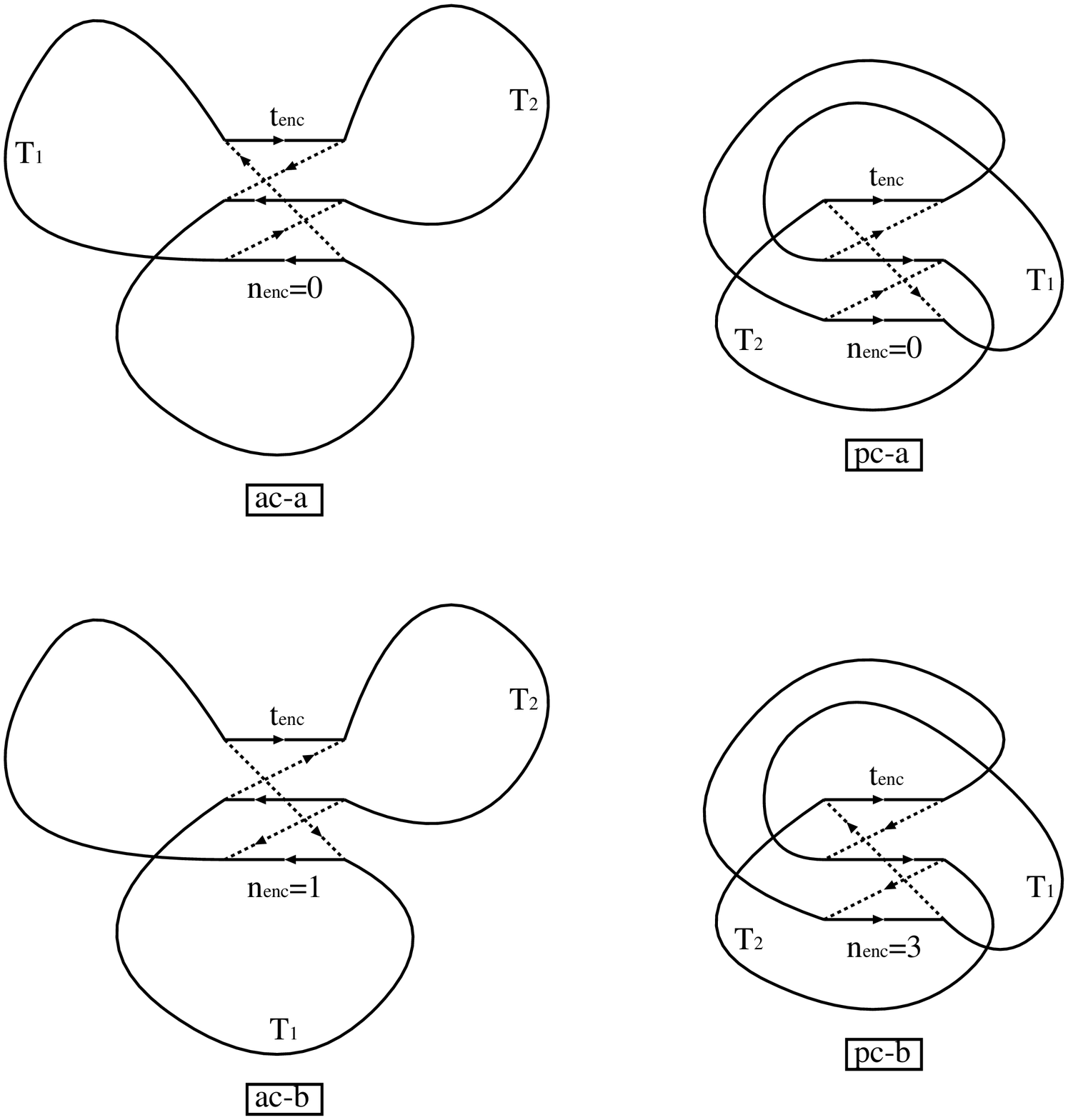}}
\caption{The periodic orbit pairs with one encounter.}
\end{figure}

\end{document}